\documentclass[useAMS,usenatbib]{mn2e}
\usepackage{graphicx}
\usepackage{hyperref}
\usepackage{natbib}
\usepackage{subfig}
\usepackage{epsfig} 
\usepackage{color}
\usepackage{ifpdf} 
\usepackage{fancyvrb}
\usepackage{txfonts}
\usepackage[all]{hypcap}

\newcommand{\dd}{\mathrm{d}}

\newcommand{\be}{\begin{equation}}                                 
\newcommand{\ee}{\end{equation}}                                   
\newcommand{\bea}{\begin{eqnarray}}                                
\newcommand{\eea}{\end{eqnarray}}                                  
                                                   
\title[QPOs from oscillating tori]{Quasi-periodic oscillations from relativistic \\ hydrodynamical slender tori}
\author[B. Mishra, F. H. Vincent, A. Manousakis, Chris P. Fragile, T. Paumard, W. Klu\'zniak]{B. Mishra$^{1}$\thanks{E-mail:
mbhupe@camk.edu.pl}, F. H. Vincent$^{1}$, A. Manousakis${^1}$, P. C. Fragile$^{2}$, T. Paumard$^{3}$, W. Klu\'zniak$^{1}$\\
$^{1}$Nicolaus Copernicus Astronomical Center, Bartycka 18, Warsaw, 00-716, Poland\\
$^{2}$Physics and Astronomy College of Charleston, 66 George Street, Charleston SC 29424, USA\\
$^{3}$LESIA, Observatoire de Paris, CNRS, UPMC, Universit\'e Paris-Diderot, 5 place Jules Janssen, 92195 Meudon, France}
\begin{document}
\date{Accepted *** Received ***}
\pagerange{\pageref{firstpage}--\pageref{lastpage}} \pubyear{2014}
\maketitle
\label{firstpage}
\begin{abstract}
We simulate an oscillating purely hydrodynamical torus with constant specific angular momentum around a Schwarzschild black hole. The goal is to search for quasi-periodic oscillations (QPOs) in the light curve of the torus. The initial torus setup is subjected to radial, vertical and diagonal (combination of radial and vertical) velocity perturbations. The hydrodynamical simulations are performed using the general relativistic magnetohydrodynamics code Cosmos++ and ray-traced using the GYOTO code. We found that a horizontal velocity perturbation triggers the radial and plus modes, while a vertical velocity perturbation triggers the vertical and X modes. The diagonal perturbation gives a combination of the modes triggered in the radial and vertical perturbations.

\end{abstract}

\section{Introduction}

Black-hole binaries, and especially black holes in low-mass X-ray binaries (LMXBs), have been extensively studied in the past decades~\citep[for a review, see][]{2004astro.ph.10551V,2006ARA&A..44...49R}.
The advent of the \textit{Rossi X-ray Timing Explorer} in the late 1990's in particular allowed the study of rapid quasi-periodic oscillations (QPOs)
in these sources, as well as in accreting neutron stars. QPOs appear as peaks (of non-zero width, hence their \textit{quasi}-periodic nature) in the power density spectra (PDS, the
simplest expression of which is the square modulus of the Fourier transform)
of some black-hole binaries. Prior to the launch of the RXTE mission, it had been suggested that QPOs may be observed in LMXBs in the kHz range as a result of inhomogeneities in the inner accretion disk about neutron stars and that they could be used to test predictions of Einstein's Theory of General Relativity (GR) in the strong field regime \citep{1990ApJ...358..538K}.
In this article, we will be only interested in the high-frequency QPOs in black hole sources, with typical frequencies of $40$ -- $450$~Hz \citep{1997ApJ...482..993M, 1999ApJ...522..397R, 2001A&A...372..551B, 2001ApJ...563..928M, 2001ApJ...552L..49S, 2001ApJ...554L.169S, 2002ApJ...564..962R, 2012ApJ...747L...4A, 2012MNRAS.426.1701B}.
These QPOs exhibit characteristic frequencies of the order of the Keplerian frequency (i.e. the orbital frequency on a circular timelike geodesic
in the equatorial plane)  
very close to the central black hole. The highest stable geodesic Keplerian frequency is reached at the
innermost stable circular orbit (ISCO).
However, in a thick accretion disk or in a torus, in principle, the orbiting fluid can exhibit quasi-stable motion at frequencies as high as that in the marginally bound orbit.
 The ISCO frequency of a $10~M_{\odot}$ black hole
is $\approx 220$~Hz, which is in the typical range for high-frequency QPOs of black-hole binaries. Consequently, it is reasonable to investigate whether
QPOs might be due to phenomena arising in the strong-field region very close to the black hole event horizon. If this picture is true, QPOs could potentially
be used to constrain the black holes properties and in particular its spin parameter~\citep{2001A&A...374L..19A} and might
be used as probes of strong-field general relativity in the relatively distant future, when precise enough data will allow for comparison of various theories of gravity~\citep[see, e.g.,][]{johannsen10}.

In order to understand the high-frequency variability, numerous models have been proposed in the last few years. However, we are still far from understanding
these phenomena. In black holes, high-frequency QPOs are associated with the so-called steep power-law spectral state of X-ray binaries~\citep{2006ARA&A..44...49R}.
This spectral state itself is not well understood. One important goal of investigations such as ours is to try to narrow down the choice between models for high-frequency QPOs
by exploring their ability to account for observations. As of today, a large diversity of models are still advocated. Blobs of matter orbiting in the inner accretion disk have been an early favourite as the cause of a quasi-periodic modulation to the X-ray flux \citep{1998ApJ...492L..59S, 1998ApJ...492L..53C, 1998ApJ...509L..37K}, but these have been ruled out \citep{2005MNRAS.357.1288B}, at least in the neutron star LMXBs, by the high degree of coherence of the oscillation (high quality factor $Q$). Although in black holes the $Q$ value is low, so blobs are not ruled out, it seems much more likely that the QPOs reflect modulations imprinted on the photon flux by intrinsic disk oscillations, whose normal modes have been studied extensively \citep{1999PhR...311..259W, 2001PASJ...53....1K,2015MNRAS.446..240M}.

Upon analyzing certain properties of twin kHz QPOs in neutron stars,
 \citet{2001AcPPB..32.3605K} proposed that they could be accounted for by a resonance between two possible motions of the accreting fluid. The resonance model predicted that high frequency QPOs in black holes should occur in pairs, as they do in neutron stars, and that the two QPO frequencies should be in the ratio of small integer numbers, e.g., 2:1 or 3:2 ratio. These predictions were borne out with the discovery of a 450 Hz QPO in the binary black hole system GRO J1655-40 \citep{2001ApJ...552L..49S} in which a 300Hz QPO had been reported previously. \citet{2001A&A...374L..19A} pointed out that the two frequencies, 450 Hz and 300 Hz, are in a 3:2 ratio, and constrained the spin of the black hole assuming a particular resonance model.
The resonance model has been used to relate the black hole spin to the frequencies expected also for other types of resonances that can occur in nearly Keplerian disks in strong gravity \citep{2005A&A...436....1T, 2013A&A...552A..10S}.
In spite of the successful predictions of the resonance model, it is important to examine other possibilities, such as that the ratio may only be approximately 3:2, and may simply reflect the ratio of two eigenfrequencies of an accretion structure.
 Recently, \citet{dexter14} advocated a model that accounts for both the steep power-law spectral state and approximately for the 3:2 frequency ratio of twin-peak QPOs.

  A torus oscillating along its own axis may cause periodic variations in its radiative flux \citep{2004ApJ...617L..45B, 2004ApJ...603L..89K, 2004ApJ...603L..93L}. Based on their analytical work \citet{2004AIPC..714..379K} also suggested that axisymmetric, up and down, motion at the vertical epicyclic frequency can also be directly observable if it occurs in the inner regions of the disk. To produce synthetic light curves and PDS, \citet{2004ApJ...617L..45B}  performed general relativistic ray-tracing of a torus undergoing simultaneous vertical and radial eigenmode oscillations with a 3:2 frequency ratio,  finding that the (higher-frequency) vertical oscillation appreciably modulates the light curve only in GR, and not in Newtonian gravity.
\citet{2004ApJ...606.1098S} considered a hot spot radiating isotropically on nearly circular equatorial orbits around a Kerr black hole and also performed general relativistic ray-tracing to produce synthetic light curves and PDS, finding that the hot spots would have to be very elongated in the azimuthal direction to suppress the (unobserved) harmonic content. The hot spot model has also been investigated by  \citep{2014MNRAS.439.1933B}. The Rossby wave instability may also lead to the appearance of QPOs in black hole binaries \citep{2006ApJ...652.1457T, 2013A&A...551A..54V}.

Geometrically thick disks or tori \citep{1978A&A....63..221A,2013LRR....16....1A} have been of great interest in studies of QPOs using analytical as well as numerical simulations. \citet{2003MNRAS.344..978R} studied high-frequency QPOs in oscillating tori in the Newtonian as well as GR framework. \citet{2005AN....326..820K, 2006CQGra..23.1689A, 2006MNRAS.369.1235B} analytically calculated linear modes in relativistic slender hydrodynamical tori and also suggested that these modes can be excited in numerical simulations. \citet{2009CQGra..26e5011S} extended these results to non-slender tori.
\citet{2013A&A...554A..57M} and \citet{2014A&A...563A.109V} studied analytic tori subject to radial, vertical, shear and expansion perturbations using relativistic ray-tracing. 

Ray-tracing of relativistic hydrodynamical simulations of an oscillating torus was performed for the first time by \citet{2006ApJ...637L.113S}, who perturbed the radial velocity to trigger the oscillations. They found multiple  high-frequency peaks in the simulated light curve, and in this paper we are going to revisit the frequency identification with particular modes which was suggested by the authors. In order to generalize this hydrodynamical work, \citet{2006ApJ...651.1031S} used relativistic ray-tracing to compute the light curves from 3D global relativistic MHD simulations of accreting tori around black hole.  However, these authors did not see any convincing evidence of QPOs in their MHD simulations.

This article aims to further develop the study of oscillating tori as a model for high-frequency QPOs and in particular twin-peak QPOs. 
To this end we perturb the 4-velocity of equilibrium tori, use relativistic hydrodynamical simulations to compute their evolution and
relativistic ray-tracing to integrate the radiative transfer equation and produce power spectra. Such computations have been
done in the past by \citet{2006ApJ...637L.113S}. The new ingredient of our work is an investigation of  various perturbations of the
equilibrium torus at various distances from the black hole to determine which oscillation modes are triggered, depending on the initial perturbation. We consider fairly slender tori (with a small cross-sectional diameter compared to its distance from the central object), whereas \citet{2006ApJ...637L.113S} considered thick tori. Unlike for thick tori, comparison of our simulations
with previous analytical studies developed in the limit of slender tori~\citep{2005AN....326..820K, 2006CQGra..23.1689A, 2006MNRAS.369.1235B} allows an unambiguous identification of the modes responsible for the peaks in the simulated PDS.
One of the main limitations of this study is the
fact that no self-consistent model is given for the origin of the perturbation of the equilibrium torus. This point may be addressed
in future work.

The article is organized as follows: In Section 2 we present the numerical method governing the general relativistic hydrodynamical (GRHD) equations to setup the torus simulation and perform relativistic ray-tracing of the results. In Section 3 we discuss the simulated PDS of oscillating tori for various initial perturbations. In Section 4 we shall give conclusions.  We use $G=1=c$ throughout the paper.

\section{QPO modes and numerical setup}    
\subsection{Eigenfrequencies of QPO modes}
The perturbation analysis in previous studies of QPOs predicted the analytical expressions of various eigenmodes we are interested in \citep{2005AN....326..820K,
2006CQGra..23.1689A, 2006MNRAS.369.1235B, 2009CQGra..26e5011S}.  These are the the radial mode ($r$), the vertical ($\theta$) mode, the X mode, the plus mode (+), and the breathing mode (b).  We shall follow the notations of \citet{2006MNRAS.369.1235B} in defining the frequencies of the eigenmodes.  These expressions are valid for a polytropic equation of state; in this work, we assume a polytropic index of $\Gamma = 5/3$.

 The lowest order eigenmode frequencies at the center of torus, $r_c$, are
\begin{equation}
\tilde{\sigma}^2_r = \tilde{\omega}^2_{R_c} = \left(1 - \frac{6}{R_c}\right),
\end{equation}
\begin{equation}
\tilde{\sigma}^2_\theta = \tilde{\omega}^2_{\theta_c}= 1,
\end{equation}
\begin{equation}
\tilde{\sigma}^2_X = \sqrt{\tilde{\omega}^2_{R_c} + \tilde{\omega}^2_{\theta_c}}
\end{equation}
\begin{equation}
\tilde{\sigma}^2_+ = \frac{1}{3}\left[ A - B^{1/2}\right]
\end{equation}
\begin{equation}
\tilde{\sigma}^2_{\textrm{b}} = \frac{1}{3}\left[ A + B^{1/2}\right]
\end{equation}
where, $\tilde{\omega}^2_{R_c}$, $\tilde{\omega}^2_{\theta_c}$ are radial and vertical epicyclic frequencies scaled by the Keplerian orbital frequency $2{\rm \pi}\nu_{\rm K}$ at the center of the torus. $A$ and $B$ are defined by,
\begin{equation}
A = 4\left(\tilde{\omega}^2_{\theta_c} + \tilde{\omega}^2_{R_c}\right) - \frac{5}{3}\tilde{\kappa}^2_{R_c},
\end{equation}
\begin{equation}
B = \left[ \left( 4\left( \tilde{\omega}^2_{\theta_c} -  \tilde{\omega}^2_{R_c} \right) + \frac{5}{3} \tilde{\kappa}^2_{R_c} \right)^2 + 4\tilde{\omega}^2_{\theta_c} \left(\omega^2_{r_c} - \tilde{\kappa}^2_{R_c} \right) \right] ,
\end{equation}
where $\tilde{\kappa_{R_c}}$ is characteristic squared frequency scaled with orbital frequency at the center of the torus \citep{2006MNRAS.369.1235B}. Table~\ref{tab:modes} shows the predicted values of these frequencies from the above expressions.
\subsection{GRHD simulation}
We numerically solve the equations of relativistic hydrodynamics derived from the following stress-energy tensor for the torus fluid 
\begin{equation}
T_{\mu\nu} = \left(\rho + \rho\epsilon + P\right)u_\mu u_\nu + Pg_{\mu\nu},
\end{equation}
where $\epsilon$ is the specific internal energy, $\rho$ is the rest-mass density, $P$ is the fluid pressure, 
$u^{\mu} = g^{\mu\nu}u_{\nu}$ is the fluid 4-velocity and $g_{\mu\nu}$
is the Schwarzschild metric tensor. We use the Kerr-Schild coordinate system in the GRHD simulations.
From the conservation of rest mass and energy-momentum, we arrive at the following conservation equations that we evolve numerically using the general relativistic magnetohydrodynamic code Cosmos++ \citep{2005ApJ...635..723A},

\begin{equation}
\partial_tD + \partial_i\left(DV^i\right) = 0,
\end{equation}
\begin{equation}
\partial_t\mathcal{E} + \partial_i\left(-\sqrt{-g}T^i_j\right) = -\sqrt{-g}T^{k}_{\lambda}\Gamma^{\lambda}_{0k},
\end{equation}
\begin{equation}
\partial_tS_j + \partial_i\left(\sqrt{-g}T^i_j\right) = \sqrt{-g}T^{k}_{\lambda}\Gamma^{\lambda}_{jk}.
\end{equation}
where $D = W\rho$ is the generalized fluid density, $W = \sqrt{-g}u^t$ is the generalized boost factor, $V^i = u^i/u^t$ is the transport velocity, $g$ is the metric determinant,
\begin{eqnarray}
\mathcal{E} = -\sqrt{-g}T^0_0 = -Wu_t\rho h - \sqrt{-g}P
\end{eqnarray}
is the total energy density, $h = 1+\epsilon+P/\rho$ is the specific enthalpy, and
\begin{equation}
S_{\nu} = \sqrt{-g}T^0_j = Wu_j\rho h 
\end{equation} 
is the covariant momentum density.
We solve the GRHD equations using the high resolution shock-capturing (HRSC) option described in \citet{2012ApJS..201....9F} and \citet{2012MNRAS.426.1928D}.

We choose the black hole mass to be
 $M = 10\,M_{\odot}$ and set up a slender 
stationary hydrodynamical torus of cross-sectional radius, $r_{\rm cross} = 0.5\,M$ around it. We accomplish this by setting the inner and center (density and pressure maximum) radii of the torus for each model, as given in Table \ref{tab:torusconfig}. In order to understand the dependence of the results on radial position of the torus we chose two central radii, first at $R_c = 10.0\,M$ and second at $R_c = 15.0\,M$. All the simulations in our study assume a constant specific angular momentum distribution, which is defined as $l = -u_{\phi}/u_t$, where $u_{\phi}$ and $u_t$ are the covariant azimuthal and time components of the 4-velocity, respectively.
We solve for the initial internal energy distribution of the torus $\epsilon(r,\theta)$. To initialize the setup we assume a polytropic equation of state $P = \rho\epsilon\left(\Gamma - 1\right) = \kappa\rho^{\Gamma}$, so the density is given by $\rho = \left[\epsilon\left(\Gamma -1\right)/\kappa\right]^{1/(\Gamma -1)}$, with $\Gamma = 5/3$. We can choose $\kappa$ to set the density (and mass) normalization of the initial torus. Once the initial setup is fixed, the gas temperature of the torus is fixed by the gravity. The torus is evolved with time using ideal gas equation of state $P = \left(\Gamma - 1\right)\rho\epsilon$.

\begin{figure*}
\centering
\hspace{-1cm}\includegraphics[width=2.2\columnwidth]{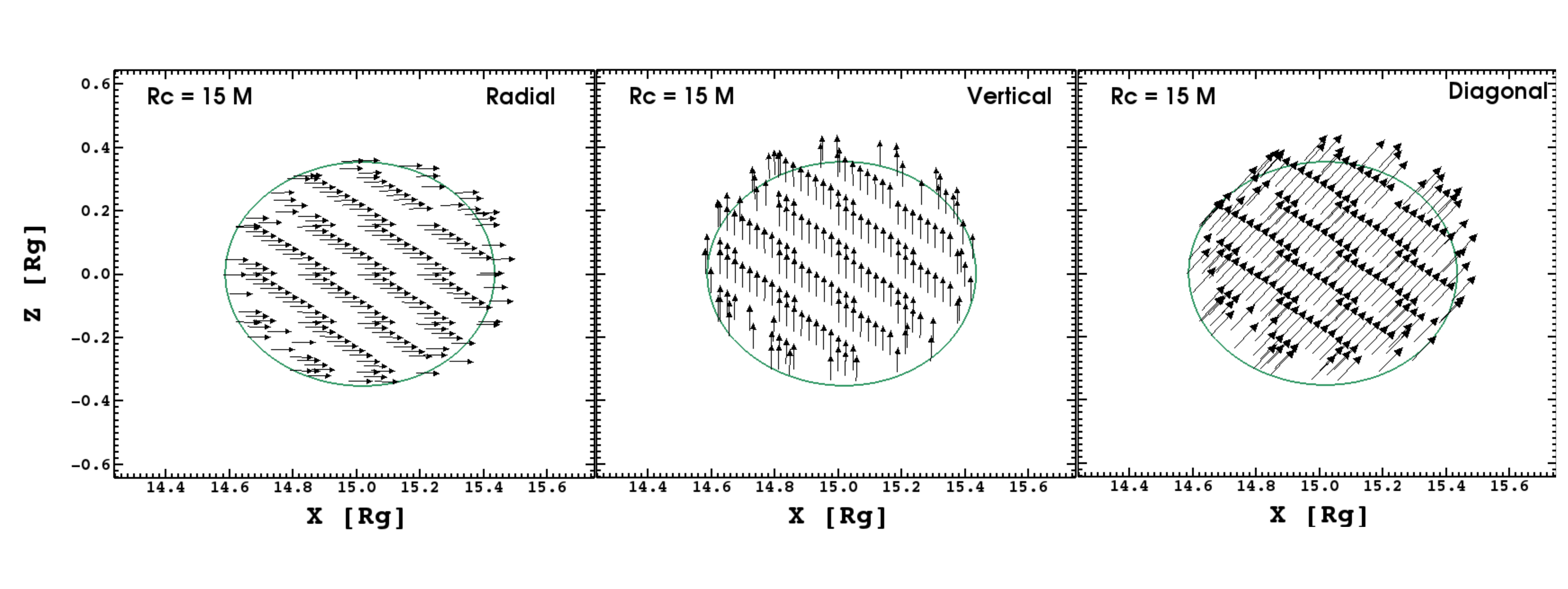}
\caption{\small{Three initial velocity perturbations to the torus. Plots only present the initial torus with pressure maximum at $R_c = 15\,M$. The torus setup at $R_c = 10\,M$ looks qualitatively similar at the corresponding radial position.}}
\label{initial}
\end{figure*}

The initial torus setup is subjected to a vertical or radial velocity perturbation using uniform velocity fields corresponding to the vertical and radial eigenmodes, respectively, in the slender torus limit \citet{2004ApJ...617L..45B,2006MNRAS.369.1235B}. The diagonal perturbation is obtained by adding these two uniform velocity fields.
 The equilibrium torus is defined by the covariant 4-velocity,
\begin{equation}
u_\mu = (u_t,0,0,u_{\phi})~.
\label{unpert}
\end{equation}
After applying the velocity perturbation, the covariant 4-velocity is given by
\be
\mathbf{u} = \left\{
\begin{array}{l}
 (u_t,\eta,0,u_{\phi}), \hspace{0.25cm}    \quad \mathrm{Radial}\\
(u_t,0,K\eta,u_{\phi}),    \quad \mathrm{Vertical} \\
(u_t,\eta,K\eta,u_{\phi}), \quad \mathrm{Diagonal} \\
\end{array}
\right.
\label{eq:Iem}
\ee
where 
\begin{equation}
K = -\sqrt{\frac{g_{\theta\theta}(R_c)}{g_{rr}(R_c)}}.
\end{equation}
This parameter is chosen such that radial and vertical 4-velocity perturbations will have the same magnitude.
The parameter $\eta$ normalizes the perturbation. We choose $\eta = 0.003 V_\mathrm{Kepler}$, where $V_{\mathrm{Kepler}} = 2\pi\,R_c \nu_c$ is the Keplerian speed at the center of torus, with $\nu_c$ being the orbit frequency.
Illustrations of the various velocity perturbations are shown in Fig.~\ref{initial}. 

The various initial setups we study are given in Table~ \ref{tab:torusconfig}. The computational domain along radial direction is defined by $r_{\mathrm{min}} = (R_{c} - 0.75)\,M$ and $r_{\mathrm{max}} = (R_c + 0.75)\,M$. The domain along the polar direction is defined by $\theta_{\mathrm{min}} = 86.4^\circ$ and $\theta_{\mathrm{max}} = 93.6^\circ$ for the torus at $R_c = 10\,M$ and $\theta_{\mathrm{min}} = 87.66^\circ$ and $\theta_{\mathrm{max}} = 92.34^\circ$ for the torus with $R_c = 15\,M$. We performed the hydrodynamics simulations with $256$ zones along both radial and polar directions. We also checked one simulation with $512\times 512$ to confirm that the results we obtained are not sensitive to the chosen resolution. All the models presented in Table~\ref{tab:torusconfig} are computed for a total coordinate time of $\Delta t = 20\,t_{\mathrm{\mathrm{orb}}}$ (orbital period at the center of the torus), where $t_{\mathrm{\mathrm{orb}}} = 198.6\,M$ for $R_c = 10\,M$ and $t_{\mathrm{\mathrm{orb}}} = 365.0\,M$ for $R_c = 15\,M$. 

\begin{table}[]
\caption{Analytically predicted frequencies (defined in Section 2.1) of the five lowest order oscillation modes of slender tori for the
two central radii considered in this study, in units of the Keplerian frequency at the
center of the torus, $\nu_c$. }
\centering                          
\begin{tabular*}{0.5\textwidth}{@{\extracolsep{\fill}}  l   c c c c c c}
\\
\hline \hline
\\
	$R_c$  & Radial  & Vertical & X & Plus & Breathing
\\
\hline \hline
\\
$10$    &  0.63 & 1 & 1.18 & 0.98 & 1.66\\
$15$      & 0.77 &1 & 1.26 & 1.18 & 1.69\\
\hline        \hline                                   
\end{tabular*}
\label{tab:modes}
\end{table}
\begin{table}[]
\caption{{Initial torus setup, the torus cross-section is always equal to $0.5\,M$. The ratio of the cross-section to the
central radius of the torus (quantifying the slenderness of the torus) is given. The last column gives the oscillation modes that dominate the power spectra (see section~\ref{sec:res}). The radial, vertical, plus and breathing modes are listed in short form as rad, vert, + and breath.
}}
\centering                          
\begin{tabular*}{0.5\textwidth}{@{\extracolsep{\fill}}  l   c c c c c c}
\\
\hline \hline
\\
Perturbation  & r$_{\rm in}$ (M)  & R$_c$ (M) &  $r_{\rm cross}/R_c$ & Observed modes
\\
\hline \hline
\\
Radial        &9.5 &10.0 & 0.05 &rad, +, breath\\ 
Radial        &14.5 &15.0 & 0.03   &rad, +, breath \\ 
Vertical      &9.5 &10.0 & 0.05  &vert, X, breath\\
Vertical      &14.5 &15.0 & 0.03 &vert, X, breath\\
Diagonal      &9.5 &10.0 & 0.05 &rad, vert, +, X, breath\\ 
Diagonal      &14.5 &15.0 & 0.03 &rad, vert, +, X, breath\\ 
\hline        \hline                                   
\end{tabular*}
\label{tab:torusconfig}
\end{table}

\subsection{Ray-tracing}
\label{sec:raytracing}
The numerical data obtained from the Cosmos++ code are ray-traced
using the open-source GYOTO\footnote{Freely available at \url{gyoto.obspm.fr}} code~\citep{2011CQGra..28v5011V}. An observer is placed at
a distance $r_{\mathrm{obs}}=10^4\,M$ from the black hole.
Photons of energy $E_{\mathrm{obs}}$ are traced backwards in time by integrating the geodesic equation
in Schwarzschild geometry.
When the photon reaches the GRHD simulation box, two different
radiation mechanisms are considered. The {\it optically thick} case assumes that blackbody
radiation is emitted at the last scattering surface of the torus. Here scattering refers
to Thomson scattering. We thus integrate the Thomson optical depth
\be
\dd \tau_{\mathrm{e}} = 0.4\, \rho\, \dd s,
\ee
where the rest-mass density $\rho$ and the proper length in the emitter's
frame $s$ are expressed in cgs units. The locus of points where $\tau_{\mathrm{e}}$ 
reaches unity is defined as the torus last scattering surface. Photons emitted
from this surface will experience no Thomson scattering and their trajectory
is thus simply defined by gravitation (they follow the null geodesics of the metric).
Blackbody radiation is emitted from this surface at the local temperature
(assuming a perfect-gas law, so that $T \propto P/\rho$) and photon energy $E_{\mathrm{em}}$ (which is
of course different from $E_{\mathrm{obs}}$ due to relativistic effects).
The {\it optically thin} case assumes that bremsstrahlung radiation is emitted
by the torus. Here, the radiative transfer equation is integrated using the
well-known emission and absorption coefficients for bremsstrahlung radiation
\citep[see e.g.][]{rybicki79}.

The GRHD simulations are scale-free in density, meaning that the mass density 
obtained from one given solution can be multiplied by an arbitrary constant factor to get
another well-defined solution. As a consequence, we have one degree of freedom
to choose which is the mass density normalization. We choose it by imposing that
the temperature at the last scattering surface be of the order of $10^7$~K, which 
fixes the density. Because of the way these two emission models are constructed, the 
optically thick case mostly traces the evolution of the torus surface (of last scattering),
while the optically thin case traces the variation of its full volume. 

For both the optically thick and thin cases, the torus
3-velocity and density are interpolated trilinearly in 2D space ($r,\theta$) and time.
When the density is evolving too abruptly between adjacent
cells to allow interpolating in this simple way (which is the case close
to the surface of the torus where the density gradient becomes quite steep), the closest neighboring grid zone is used.
The torus is assumed not to be varying in the $\varphi$ direction, but can easily be made 3D from the 2.5D simulations, since Cosmos++ evolves all three components of the spatial velocity.

\section{Power spectra of the oscillating tori}
\label{sec:res}
\begin{figure*}
\centering
\includegraphics[width=2\columnwidth]{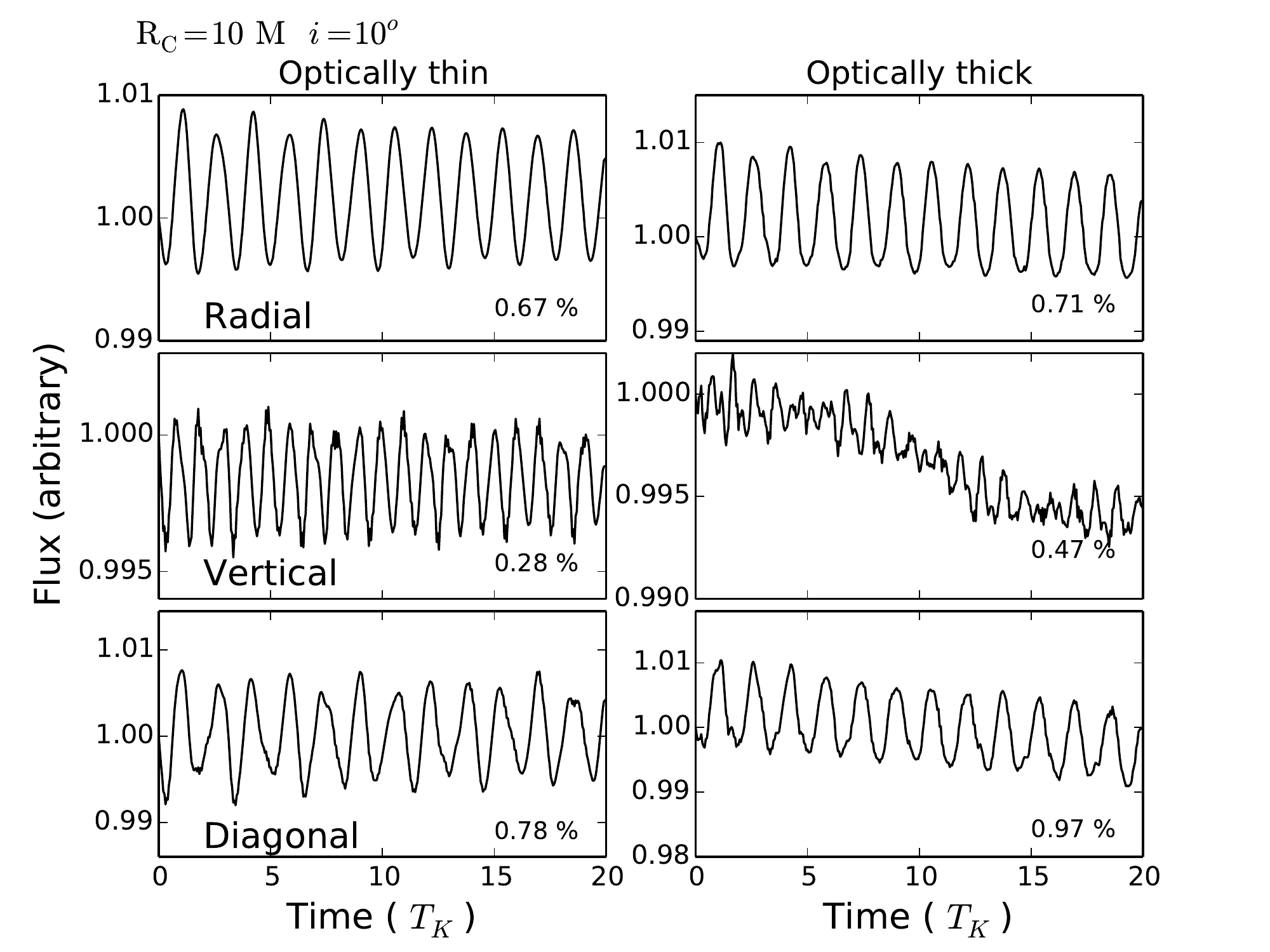}
\caption{\small{Light curves of the oscillating torus with $R_c=10\,M$,
assuming an optically thin (left) or optically thick (right) emission, for three initial velocity perturbations, radial (top), vertical (middle) and diagonal (bottom). The percentage values in each panel correspond to the pulsed fraction, i.e. the maximum relative variation of the flux.}}
\label{lc10}
\end{figure*}
\begin{figure*}
\centering
\includegraphics[width=2\columnwidth]{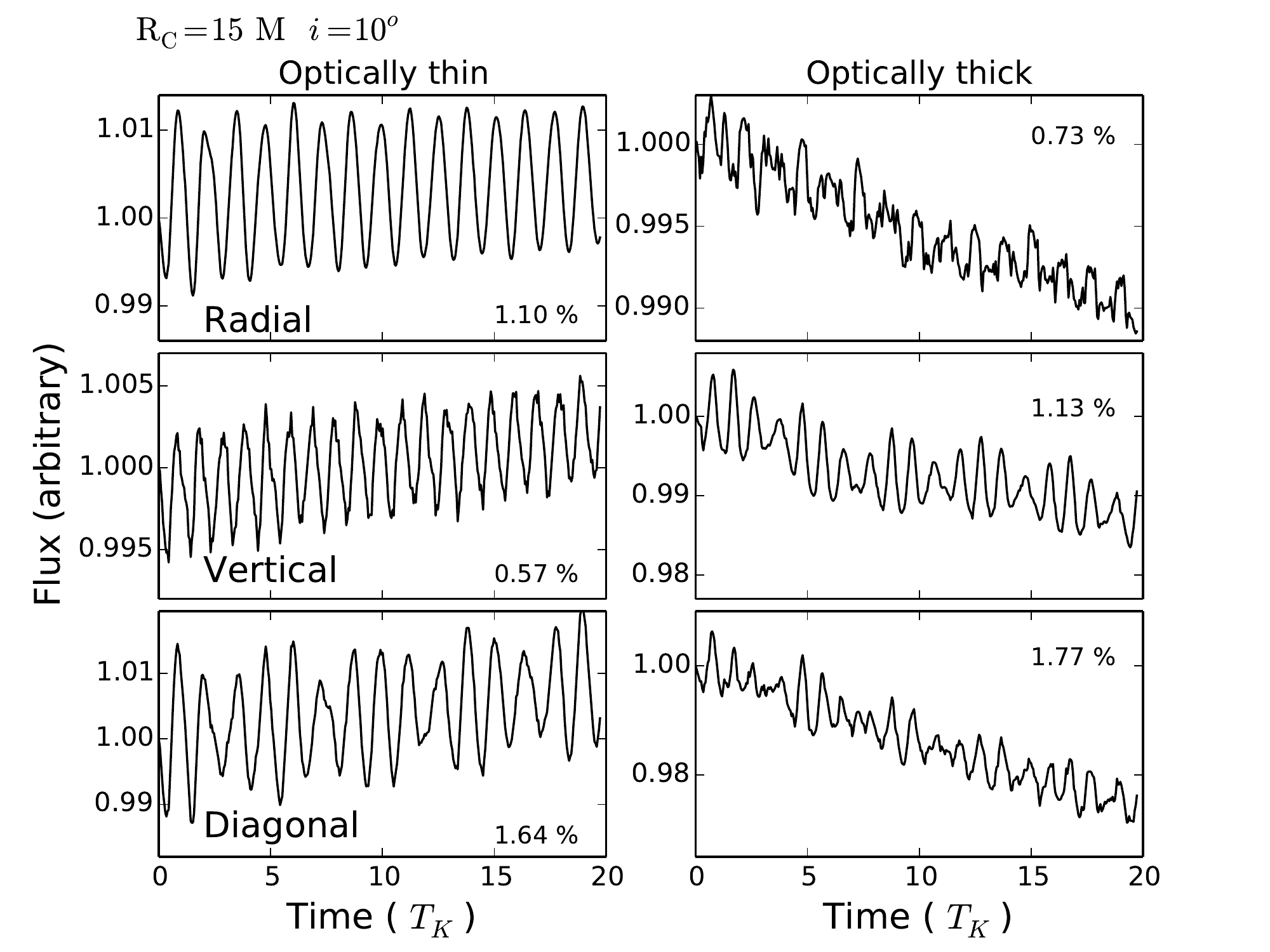}
\caption{\small{Same as Fig.~\ref{lc10}  with $R_c = 15\,M$}}
\label{lc15}
\end{figure*}
\begin{figure*}
\centering
\includegraphics[width=2\columnwidth]{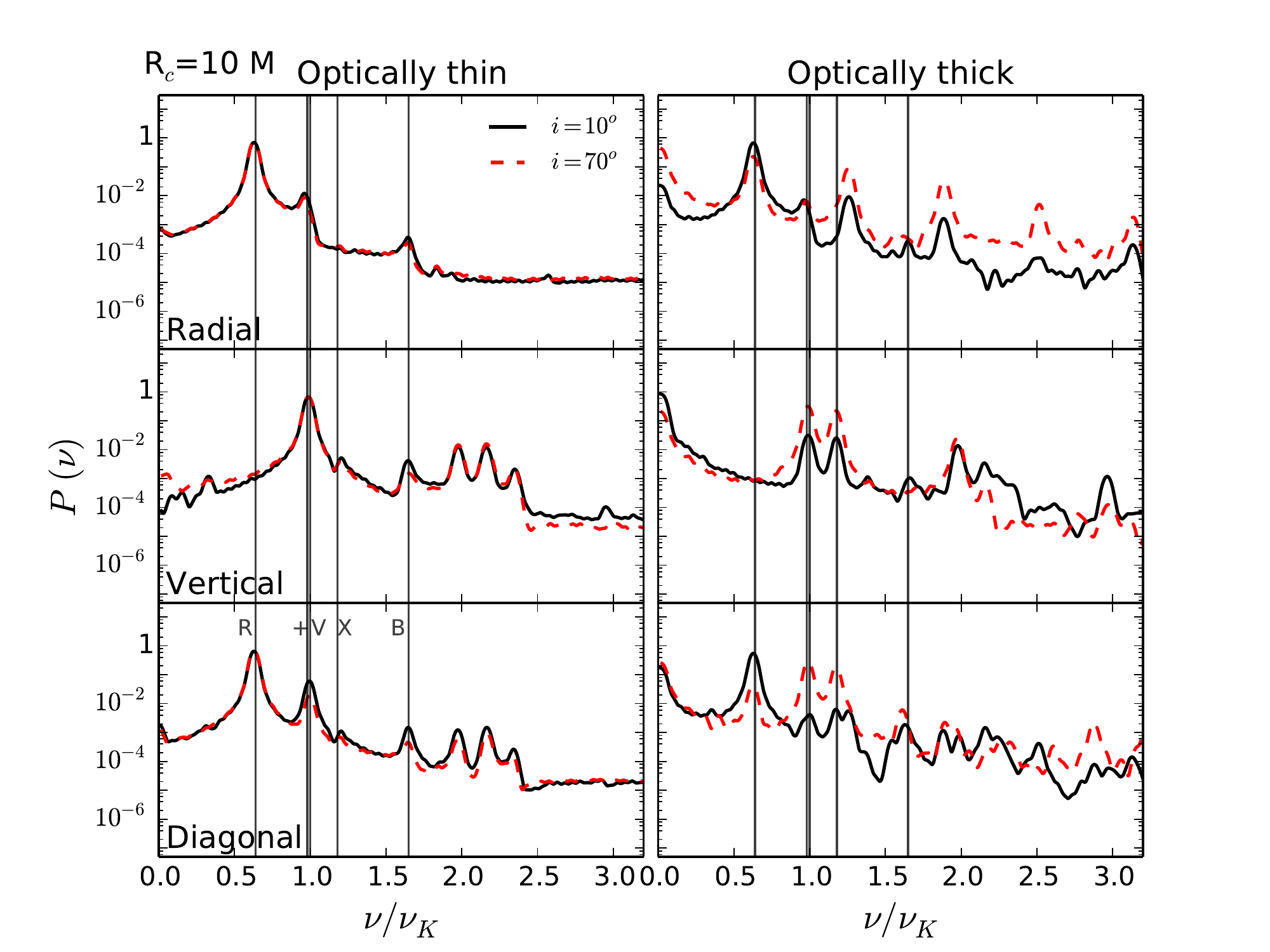}
\caption{\small{Power density spectra (PDS) of the oscillating torus with radial, vertical and diagonal velocity perturbation (top to bottom). The initial pressure maximum of the torus lies at $R_c = 10\,M$. The horizontal axis corresponds to frequency normalized to the Keplerian frequency at the center of the torus, $\nu_c$. The solid black and dashed red curves show the PDS for two inclinations $i = 10^\circ$ and $i = 70^\circ$, respectively. The left and right panel correspond to optically thin and thick cases, respectively. The vertical gray lines are located at the values of frequencies of the five lowest order oscillation modes for a $R_c=10\,M$ torus, as given in Table~\ref{tab:modes}.}}
\label{all_rc10}
\end{figure*}
\begin{figure*}
\centering
\includegraphics[width=2\columnwidth]{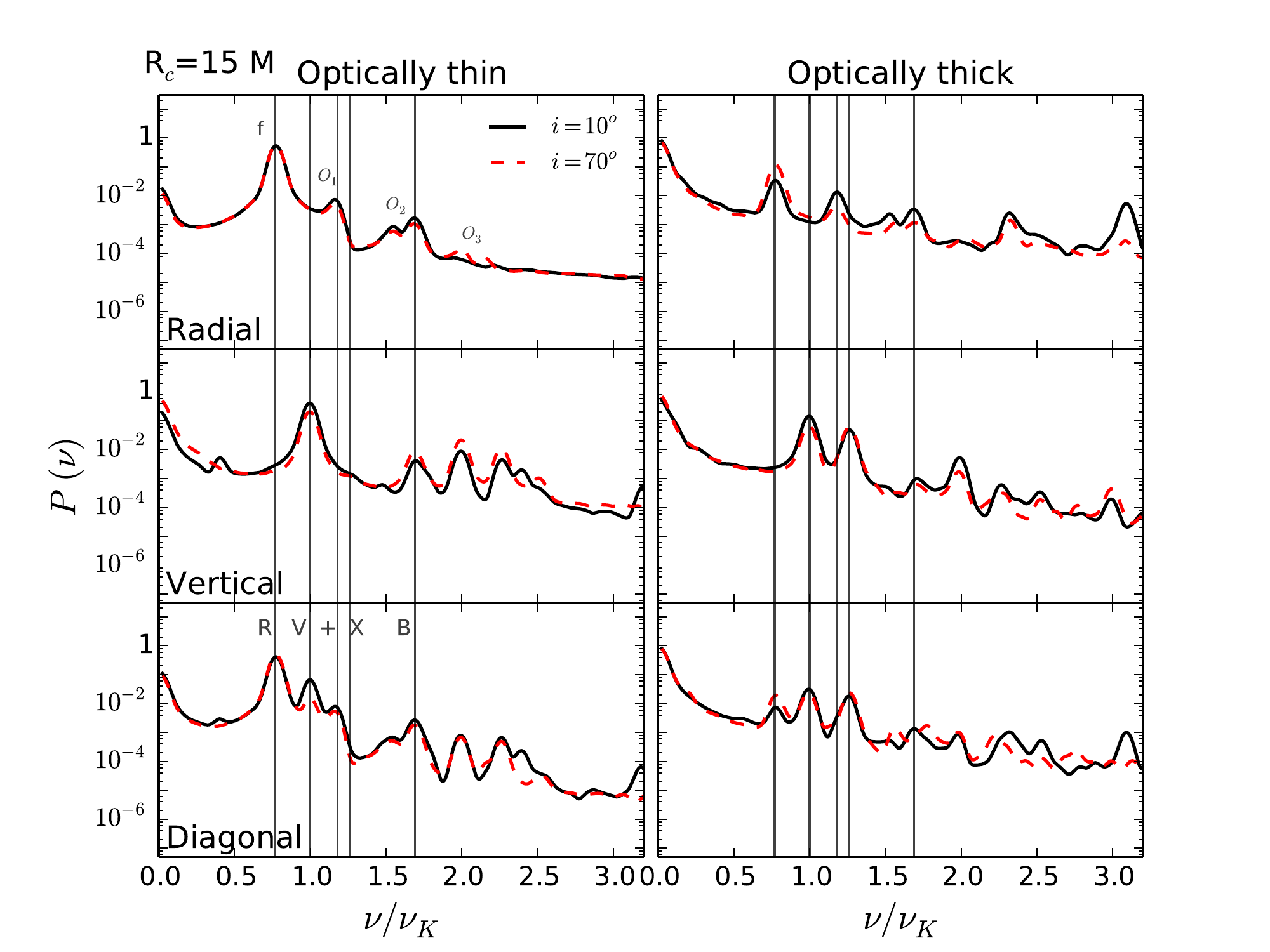}
\caption{\small{Same as Fig.~\ref{all_rc10} with $R_c = 15\,M$ torus.}}
\label{all_rc15}
\end{figure*}

The light curves from each of our simulations and each of our radiation models are shown in Figs.~\ref{lc10} and \ref{lc15}, for an observer inclination of $i=10^{\circ}$ (nearly face-on view). For all cases, the maximum relative flux variation (pulsed fraction) is of the order of $1 \%$. This is in reasonable agreement with observations. In Fig.~\ref{lc15} we also notice a small increasing trend in the optically thin lightcurves and decreasing trend in optically thick ones.  Both are due to numerical fluctuations of density from cell to cell in the hydrodynamical simulations.  

The Lomb-Scargle PDS \citep{lomb1989} of all light curves (two radial positions of torus with two different inclinations) are shown in Figs.~\ref{all_rc10} and~\ref{all_rc15}. The first obvious feature appearing from the PDS is that all of them are dominated by peaks corresponding to the mode(s) of initial perturbation. More interesting is the presence of additional peaks in all PDS. Given that the diagonal perturbation PDS appear simply as the sums of the radial and vertical PDS, we shall mainly focus on these last ones in our discussion.

In the {\it radial} PDS of Figs.~\ref{all_rc10} and \ref{all_rc15}, the prominent peaks correspond to the radial, plus, and breathing modes, as well as the sum of the radial and plus modes and higher harmonics of the radial mode. We note that the radial, optically thin PDS at $R_c=15\,M$
is comparable to the results obtained by \citet{2006ApJ...637L.113S} for the same radial position of the center of torus, although our torus
is much more slender (about a factor of 100). The peaks in \citet{2006ApJ...637L.113S} with labels $f$, $o_1$, $o_2$, $o_3$ (reproduced, for convenience, in the upper panel of our Fig.~\ref{all_rc15}) correspond to our radial (57 Hz), plus (82 Hz), twice the radial (114 Hz) and sum of radial and plus modes (140 Hz).  
In addition we also see a breathing mode, which was not present in \citet{2006ApJ...637L.113S}. We argue that this is due to the effect of the background on the slender torus. If we increase the cross section of our torus and make it more like \citet{2006ApJ...637L.113S}, the background does not affect the torus significantly and we do not see the breathing mode. 

In the {\it vertical} PDS in Fig.~\ref{all_rc10}, the vertical and breathing modes are present in both optically thin and thick cases. In addition, the optically thick case shows a prominent peak at the X-mode frequency, though it is much less prominent in the optically thin case. This is due to the fact that in the optically thick case we are tracing the surface and in the optically thin case the volume. The X-mode causes more fluctuations on surface than in volume. Fig.~\ref{all_rc15} has similar features as Fig.~\ref{all_rc10} except that we almost do not see the X-mode in this case. 
Three higher-frequency harmonics and combinations
of frequencies also appear with strong power in both tori. 
They correspond to two times the vertical
frequency, two times the X-mode frequency, as well as the sum of the vertical and
X-mode frequencies. We note that multiplying two sinusoidal functions
gives rise to a signal varying at the sum of their frequencies. 
The ``vertical + X'' peak could thus
be interpreted as a non-linear coupling between these two modes.
However, it seems more likely that it is related to the initialization of the vertical velocity perturbations.  In all our PDS except for the optically-thick, vertically-perturbed case, there are clearly dominant peaks corresponding to the initial velocity perturbation mode, with all other oscillation modes appearing significantly weaker.  Only in the case of the optically-thick, vertically-perturbed case is one of these ``secondary'' peaks of comparable strength.  However, we note that the velocity eigenvector field for a vertically-oscillating, non-slender, torus contains within it the velocity field of an X-mode ~\citep[see Fig.~5 of][]{2007ApJ...665..642B}. Therefore, it seems most likely that this prominent X-mode is also the result of the initial perturbations and not evidence of non-linear coupling.

We also propose an intuitive explanation of the fact that a radial perturbation of the torus
triggers the plus mode, whereas a vertical perturbation
triggers the X-mode (qualitative sketch in Fig.~\ref{coupling}). When the torus is excited vertically, it will essentially
oscillate along the $z$-axis at the {\it local}, vertical epicyclic frequency.  Likewise, when it is excited radially
it will oscillate in the $r$-direction at the {\it local}, radial epicyclic frequency. These frequencies
are functions of the coordinate radius $r$; thus different parts of the torus will
oscillate at slightly different frequencies: the parts of the torus closest to the
black hole will oscillate at a frequency slightly larger than the more distant parts.
As a consequence, the torus cross-section will be distorted, roughly following 
the X mode pattern in the vertical case, and the plus mode pattern in the radial
case. All PDSs in our study show precisely this trend.  

Another observation that come from our data is that the breathing
mode, present in all simulations, is stronger when the torus is further from
the black hole. We believe that this is an anomaly owing to the fact that the
background (which has the same characteristics in all simulations) acts more strongly
on the more slender torus. Given that its cross-sectional radius is constant, the torus becomes relatively
more slender when it is displaced away from the black hole. This bigger interaction of
the background with the torus leads to a more prominent breathing mode. We also 
believe that the presence of the breathing just arises due to the initial mismatch between
the torus and the background. This kind of initialization issue has been reported even in studies of unperturbed tori \citep{2007A&A...467..641S}.

We do not note a strong inclination dependence in our optically thin PDS, which seems
contradictory to previous works, in particular \citet{2004ApJ...617L..45B}.
However, these findings have been obtained for an incompressible torus
and the inclination dependence is a very strong function of the torus
equation of state (M. Bursa, private communication). The inclination
dependence reported in~\citet{2006ApJ...637L.113S} is also obtained for
a different equation of state ($\Gamma=4/3$), and for a much thicker torus.
Only a full parameter study, which goes beyond the scope of the present
paper and will be the subject of future work, will allow a more thorough exploration of the inclination dependence.

To recapitulate, the most important conclusion of our analysis is that the
dominant modes appearing in the PDS are the radial, plus and breathing
modes for a radial perturbation, and the vertical, X and breathing modes for
a vertical perturbation. However, we do not observe a clear domination of peaks in a 
3:2 ratio (X/vertical, plus/radial, or vertical/breathing). Our current results thus do not
show any natural selection mechanism for the 3:2 QPOs. However, the dominant
peaks might be (and most probably are) dependent on the initial condition (in particular
the initial velocity field and the torus size). We will devote future study to determining if some configurations naturally generate dominant 3:2 oscillations.

\section{Conclusions}
\begin{figure}
\centering
\includegraphics[width=1\columnwidth]{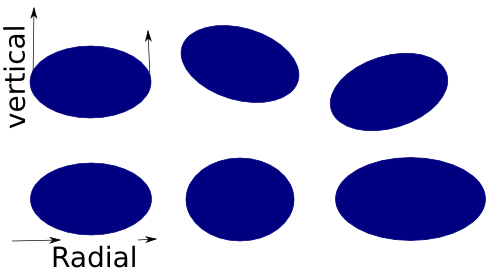}
\caption{\small{Qualitative illustration of the response of a torus to a uniform vertical (top) and radial (bottom) 4-velocity perturbations.}}
\label{coupling}
\end{figure}

We performed hydrodynamic simulations of oscillating slender tori and ray-traced the results to obtain power density spectra. We concentrated on three types of perturbations: radial, vertical and diagonal. The important conclusions of the article concern the dominant modes, possibility of presence of non-linear coupling of vertical and X-mode, an absence of inclination dependence in optically thin case and the different distributions of mode power for different radiative transfer models. To summarize:
\begin{itemize}
\item[1.] We found that a uniform radial perturbation of the velocity filed can trigger both the radial mode and the plus mode. Likewise, a uniform vertical perturbation of the velocity field can trigger the vertical mode and an X-mode.  This point is illustrated in Fig.~\ref{coupling}.
We also found that no matter where we place the center of the torus, we see a similar response. We argue that the presence of both the radial and plus mode or the vertical and X-mode is owing to non-uniform velocity vectors of the vertical and radial eigenmodes in an even slightly non-slender torus. 
\item[2.] We did not find any clear dependence on inclination in our ray-traced light curves for optically thin case, whether the torus was nearly edge-on or nearly face-on. We suspect that this is due to the fact that our tori are very slender and compressible. 
\item[3.] We found that the two different radiative models we considered can give very different lightcurves and hence different PDS. In particular, the optically thin and thick cases give different powers in the plus- and X-modes. 

 \end{itemize}
 
\section*{Acknowledgments}
We thank Ji\v ri Horak, Omer Blaes, Michal Bursa and Odele Straub for enlightening 
discussions at various stages of this work. MB also thank to College of Charleston SC for hosting during final stage of project. Research was supported by Polish NCN grant UMO-34 2011/01/B/ST9/05439 and 2013/08/A/ST9/00795.
Part of the work was supported by the Polish NCN grant 2013/09/B/ST9/00060.
\bibliographystyle{mn2e_fixed}
\bibliography{ref}
\end{document}